%% file: root.tex
\documentclass{ifacconf}

\usepackage{graphicx}      % include this line if your document contains figures
\usepackage{natbib}        % required for bibliography
\usepackage{siunitx}
\usepackage{color}
\usepackage[official]{eurosym}
\usepackage{todo}
\usepackage{amsmath}
\usepackage{bm}
\usepackage[absolute,verbose]{textpos}
\begin{document}
\begin{frontmatter}

\title{EduBal: An open balancing robot platform for teaching control and system theory} 
% Title, preferably not more than 10 words, oh yeah.

\author[First]{Christian-Eike Framing} 
\author[Second]{Raffael Hedinger} 
\author[Third]{Emmanuel Santiago Iglesias}
\author[First]{Frank-Josef Heßeler}
\author[First]{Dirk Abel}

\address[First]{Institute of Automatic Control, RWTH Aachen University, DE-52074 Aachen, Germany (e-mail: \{c.framing\}\{f.hesseler\}\{d.abel\}@irt.rwth-aachen.de).}
\address[Second]{Institute for Dynamic Systems and Control, ETH Zürich, 8092 Zürich, Switzerland (e-mail: hraffael@idsc.mavt.ethz.ch)}
\address[Third]{RWTH Aachen University, DE-52074 Aachen, Germany (e-mail: emmanuel.iglesias@rwth-aachen.de)}

\begin{abstract}                % Abstract of not more than 250 words.
In this work we present EduBal, an educational open-source hardware and software platform for a balancing robot. The robot is designed to be low-cost, safe and easy to use by students for control education. Along with the robot we present example tasks from system identification as well as SISO and MIMO control. Using Simulink, students can quickly implement their control algorithms on the robot. Individual control parameters can be tuned online while analyzing the resulting behavior in live signal plots. At RWTH Aachen University and ETH Zurich 28 units have so far been built and used in control classes. In first laboratory experiences students show high intrinsic motivation and creativity to apply the studied concepts of control theory to the real system.
\end{abstract}


\end{frontmatter}
%===============================================================================
\begin{textblock*}{\textwidth}(13.85mm,30mm)
\small\textcopyright 2020 the authors. This work has been accepted to IFAC for publication under a Creative Commons Licence CC-BY-NC-ND
\end{textblock*}
\section{Introduction}
Control theory, while being essential for many of today's innovations, can seem an abstract and unfamiliar topic for students. Employing interactive learning as part of the curriculum, especially on a real system, can build a bridge between theory and practice. As part of it, students may develop an intuitive understanding of individual control characteristics.
\begin{figure}[t]
\graphicspath{{img/}}
\def\svgwidth{\columnwidth}
{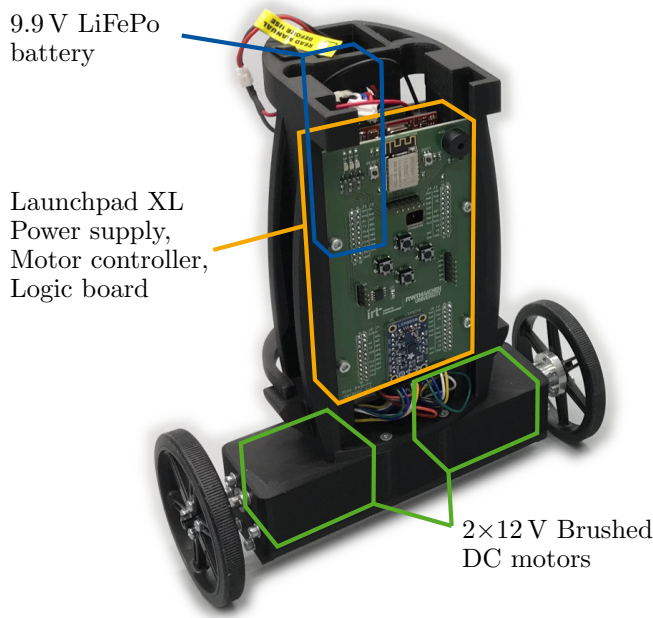}

\caption{Front view of EduBal with its individual parts highlighted}
\label{fig:edubal_overview}
\end{figure}
In the past, low-cost educational systems such as a mobile robot \citep{mondadaEpuckRobotDesigned2009} and a robotic arm \citep{sorianoLowCostPlatform2014} have been proposed. A balancing robot, like the inverted pendulum, is an unstable, nonlinear and multiple-input-multiple-output (MIMO) system allowing a wide range of control-theoretic applications. Balancing robot designs for education have appeared in the literature \citep{howardEnhancingLinearSystem2015,gonzalezLowCostTwowheels2017} and are commercially available\footnote{e.g.\ the Balboa Balancing Robot by Pololu and the GoPiGo3 BalanceBot by Dexter Industries}. Most of these designs are built around the inexpensive Arduino microcontroller\footnote{with the exception of the Raspberry Pi in the GoPiGo3}. This microcontroller is well supported by MATLAB/Simulink but has only limited processing power for useful features such as live communication or advanced control algorithms like Model Predictive Control. Some of the hardware designs may also be too fragile for direct student use.

In this work we present the \emph{Edu}cational \emph{Bal}ancing Robot \emph{EduBal}. It is a low-cost and robust design enabling the safe experimentation with a wide range of control-theoretic aspects from system analysis, control design and system identification. The selected microcontroller and actuators offer the potential for demanding control algorithms. For a wider audience, its hardware and software design is made openly available in the project repository\footnote{https://gitlab.com/rwth-irt-public/edubalembedded} under the permissive MIT license. 

In the remainder of the paper, we first derive a physical model of the robot in Section \ref{sec:model}. The model is used to demonstrate example student tasks from SISO and MIMO control design in Section \ref{sec:control_design}. Section \ref{sec:implementation} gives details about the specific hardware, software and safety implementation of the robot. First experiences from educational use are described in Section \ref{sec:educational_use} before a conclusion is made and future work is suggested in Section \ref{sec:conclusion}.
\section{Robot model}
\label{sec:model}
\begin{figure}
\centering
\graphicspath{{img/}}
\def\svgwidth{0.5\columnwidth}
{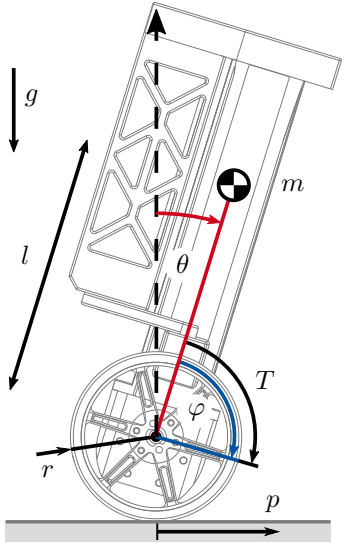}

\caption{Lateral plane view of EduBal}
\label{fig:edubal_definitions}
\end{figure}
Depicted in Fig. \ref{fig:edubal_definitions} is a lateral plane view of an EduBal model. The robot has two wheels of radius $r$ which are assumed weightless. The wheel rotation angle $\varphi$ is measured relative to the body axis and leads to the longitudinal position $p$. The upper body has mass $m$ which is assumed to be located at a distance $l$ from the wheel axle. Due to gravity $g$ acting on the body, it may pitch with an angle $\theta$. The electric motors, which are fixed to the upper body, produce a combined torque $T$ acting on both the wheel and body. The experimentally determined model parameters are given in Table \ref{tbl:parameters_table}.
\input{img/parameters_table}

\subsection{Mechanics}
The mechanical dynamics of the robot can be derived using the Lagrange formalism \citep{frankovskyModelingTwoWheeledSelfBalancing2017}:
\begin{equation}
\frac{d}{dt}\left(\frac{\partial L}{\partial\dot{q}_i}\right)-\frac{\partial L}{\partial q_i} = Q_i - \frac{\partial D}{\partial \dot{q}_i}
\end{equation}
where the Lagrangian is defined as the difference between the kinetic and potential energy of body and wheels:
\begin{equation}
L = E_k - E_p = E_{kB} + 2E_{kW} - E_{pB} - 2E_{pW}
\end{equation}
The generalized forces acting on the system are the motor torques $Q_i=T$ and the generalized coordinates $q_i=\left(p,\theta\right)^T$. We neglect friction on both the wheel axle and ground and assume weightless wheels. Applying the Lagrangian formalism yields the two differential equations:
\begin{eqnarray}
rlm\ddot{\theta}\cos(\theta)+rm\ddot{p}-rlm\sin(\theta)\dot{\theta}^2 &= T,\label{eq:nonlin1}\\
2l^2m\ddot{\theta}+lm\ddot{p}\cos(\theta) - glm \sin(\theta) &= -T.\label{eq:nonlin2}
\end{eqnarray}

For further analysis we linearize the system about the upper equilibrium point using the approximations:
\begin{equation}
\cos(\theta)\approx 1, \qquad \sin(\theta)\approx \theta, \qquad \dot{\theta}^2\approx 0\label{lin_approximations}
\end{equation}
Substituting the terms \eqref{lin_approximations} in \eqref{eq:nonlin1} and \eqref{eq:nonlin2} as well as equating both equations gives the linearization:
\begin{equation}
rlm\ddot{\theta} + rm\ddot{p} = -2l^2 m \ddot{\theta}-lm\ddot{p} + glm\theta\label{eq:lin_combined}
\end{equation}
\subsection{Electric motor}
\label{sec:electric_motor}
The torque $T$ is generated by the electric motors. For stabilization control design, we initially neglected their internal dynamics and modeled them as simple proportional element. During controller validation however, the motor gearbox showed significant backlash. This nonlinearity could not directly be incorporated into the linear model and complicated stabilization control. In order to mitigate this effect, we designed an underlying feedback proportional-integral controller (PI) for the wheel velocity~$\dot{\varphi}$. 

The PI controller was designed using pole-placement. An often used representation of the electric motor is a linear second-order model describing the electrical and mechanical dynamics. The corresponding motor parameters may be estimated using a manual armature resistance measurement and least-squares identification on measurement data \citep{saabParameterIdentificationDC2001}. With the goal of stabilization control, it is however sufficient to identify only the slower mechanical dynamics as a first-order model. In an experiment a voltage step sequence $U$ was applied to the electric motor and the resulting wheel velocity $\omega=\dot{\varphi}$ was measured. The following dynamic model could be identified:
\begin{equation}
G_{\mathrm{EM}}=\frac{\omega(s)}{U(s)}=\frac{2.6}{0.038s + 1}
\end{equation}

The transfer function of the closed-loop system has two poles and one zero. The poles were placed so that a relatively long settling time of $t_s=\SI{0.5}{\s}$ for the dominant pole, and a natural frequency of at most $f=\SI{30}{\per\s}$ were achieved. The goal was to reduce the actuator bandwidth and in consequence decrease excitation of the critical high frequencies. 

For system analysis and control design, this underlying control loop needs to be included in the model. We approximate the closed-loop dynamics using a first order lag element
\begin{equation}
\ddot{\varphi} = -\frac{1}{t_{\mathrm{EM}}}\dot{\varphi} + \frac{K}{t_{\mathrm{EM}}} \dot{\varphi}_{\mathrm{ref}}\label{eq:em_pt1}
\end{equation}
where $\dot{\varphi}_{\mathrm{ref}}$ is the desired motor speed, $K$ is the gain and $t_{\mathrm{EM}}$ the time constant. The given parametrization leads to $K=1$ and $t_{\mathrm{EM}}=\SI{0.0994}{\s}$. 

\subsection{Combined system model}
By solving \eqref{eq:lin_combined} for $\ddot{\theta}$ and using \eqref{eq:em_pt1} with the relation $p=r(\varphi+\theta)$ we formulate the overall robot model as a linear state-space system
\begin{equation}
\begin{aligned}
\bm{\dot{x}} &= \bm{Ax+Bu}\\
\bm{y} &= \bm{Cx}
\end{aligned}\label{eq:state_space}
\end{equation}
with state vector $\bm{x} = (\varphi, \dot{\varphi}, \theta, \dot{\theta})^T$, input $u = \dot{\varphi}_{\mathrm{ref}}$ and output $\bm{y} = \left(p, \theta\right)^T$. The system matrices result as
\begin{eqnarray}
\begin{aligned}
\bm{A} &= \left(\begin{array}{cccc}
0 & 1 & 0 & 0\\
0 & -\frac{1}{t_{\mathrm{EM}}} & 0 & 0\\
0 & 0 & 0 & 1\\
0 & \frac{lm+rm}{t_{\mathrm{EM}} \eta} & \frac{glm}{r\eta} & 0
\end{array}\right)
\qquad
\bm{B} = \left(\begin{array}{c}
0\\
\frac{K}{t_{\mathrm{EM}}}\\
0\\
-\frac{l m + rm}{t_{\mathrm{EM}} \eta}
\end{array}\right)\\
\bm{C} &= \left(\begin{array}{cccc}
r & 0 & r & 0\\
0 & 0 & 1 & 0
\end{array}\right)
\end{aligned}
\end{eqnarray}
where
\begin{equation}
\begin{aligned}
\eta &= 2ml + mr + \frac{2ml^2}{r}.
\end{aligned}
\end{equation}
The matrix $\bm{A}$ has eigenvalues with positive real part $\text{Re}(s_i)>0$, hence the system is unstable for the upper equilibrium point.
As the controllability matrix
\begin{equation}
\bm{Co} = \left(
\begin{array}{cccc}
\bm{B} & \bm{AB} & \bm{A}^2 \bm{B} & \bm{A}^3 \bm{B}
\end{array}
\right)
\end{equation}
has full rank, the system is controllable and a stabilizing feedback controller can be designed.
\section{System analysis and Control design}
\label{sec:control_design}
\subsection{SISO Position control}
\label{sec:siso_pos_control}
\begin{figure}
\graphicspath{{img/}}
{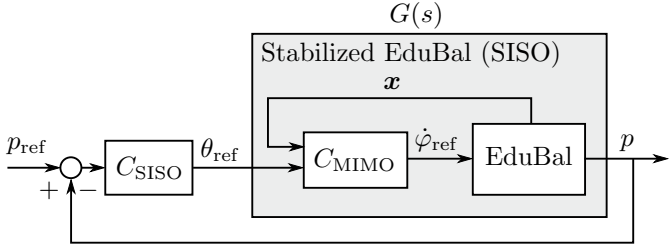}
\caption{Position control of robot as SISO problem}
\label{fig:edubal_siso}
\end{figure}
One motivation for designing EduBal was the use of it as a demonstrative example in basic control theory. Most of the relevant methods such as pole-zero analysis and root-locus control design apply for SISO systems. The task of stabilizing the full robot around its equilibrium point however is a MIMO control problem which is treated in more advanced control theory classes. For SISO stability analysis and control design the problem of controlling the robot's position can be used (see Fig. \ref{fig:edubal_siso}). 

\begin{figure}
\includegraphics[scale=1]{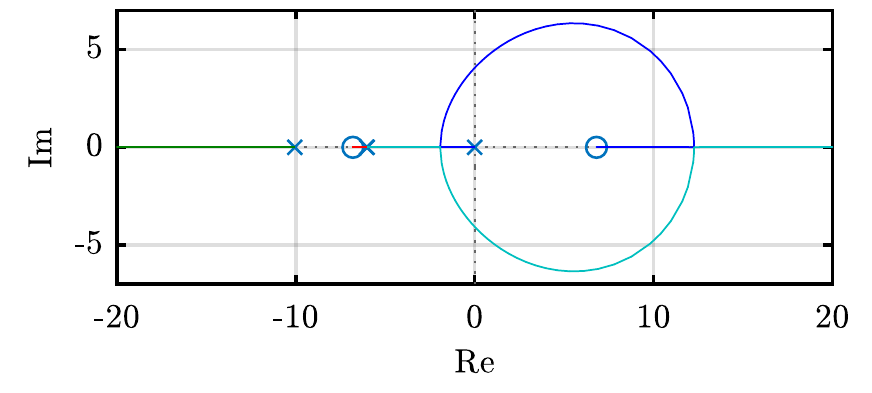}
\caption{Root locus diagram of $G(s)$ in Fig. \ref{fig:edubal_siso}}
\label{fig:edubal_pos_control_polezero}
\end{figure}

A controller $C_{\mathrm{MIMO}}$ that stabilizes the body angle needs to be designed beforehand. We use a three-state linear quadratic regulator (LQR) that controls the state $\tilde{\bm{x}}$:
\begin{equation}
\tilde{\bm{x}}=\left(
\begin{array}{c}
\dot{\varphi}\\
\theta-\theta_{\mathrm{ref}}\\
\dot{\theta}
\end{array}
\right)
\end{equation}
Using $\theta_{\mathrm{ref}}$, a target pitch angle $\theta_{\mathrm{ref}}\neq 0$ can be achieved. The controller gain $\bm{K}_{\mathrm{LQR},3\times 3}$ is determined by minimizing the quadratic cost function
\begin{equation}
J(u)=\int_{0}^{\infty} (\bm{x}^T\bm{Q}\bm{x} + uRu)dt\label{eq:lqr_cost}
\end{equation}
where $\bm{Q}=\text{diag}\left(
0, 10^0, 0\right)$ and $R=10^2$ penalize control error in body angle and actuation energy respectively. Numerically solving the resulting Riccati equation gives the optimal controller parameters in continuous time:
\begin{equation}
\bm{K}_{\mathrm{LQR},3\times 3} = \left(-2, -88.75, -14.73\right)
\end{equation}
The resulting closed-loop system with $C_{\mathrm{MIMO}}$ is SISO and has the input $\theta_{\mathrm{ref}}$ and the output $p$. It is provided to the students in the form of a transfer function:%ToDo: Raffael fragen, ob Werte sich geändert haben
\begin{equation}
\begin{split}
G(s)&=\frac{P(s)}{\theta_{\mathrm{ref}}(s)}=\frac{-27.96 s^2 + 1297}{s^4 + 22.11 s^3 + 157.6 s^2 +365.3s}
\end{split}
\end{equation}
The characteristics of the stabilized robot can be analyzed using the pole-zero diagram (refer to the root-locus diagram in Fig. \ref{fig:edubal_pos_control_polezero}). Most notably it includes an integrator and a zero which does not have minimum phase. Within the laboratory, students are asked to link these characteristics to the observed behavior of the robot. As $C_{\mathrm{MIMO}}$ does not control the wheel angle $\varphi$, a body angle reference $\theta_{\mathrm{ref}}\neq 0$ will lead to a position drift $p$ over time which corresponds to integrating behavior. The non-phase-minimum zero characteristic may be observed as an undershoot in the step response of the system. Intuitively, in order to lean forward ($\theta>0$), the robot briefly needs to drive in reverse ($\dot{p}<0$) before changing direction ($\dot{p}>0$) to keep body angle constant.

\begin{figure}
\includegraphics[scale=1]{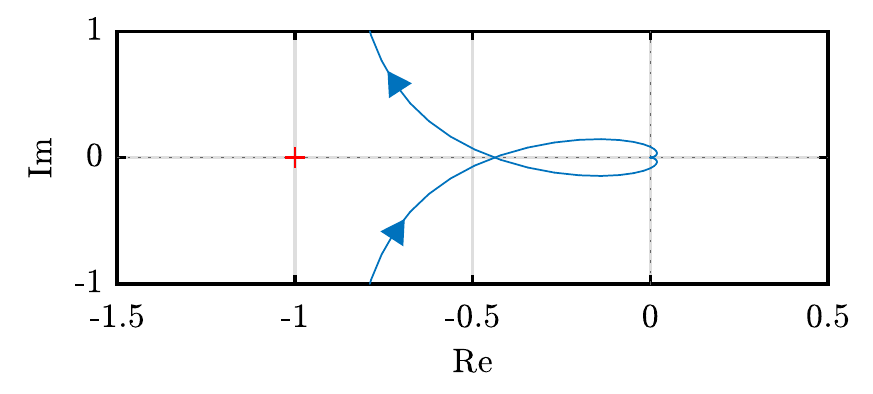}
\caption{Nyquist diagram of $G(s)$ in Fig. \ref{fig:edubal_siso}, here for a stable parametrization $K_p=0.58$.}
\label{fig:edubal_siso_nyquist}
\end{figure}
Students are then tasked to design a controller $C_{\mathrm{SISO}}$ which stabilizes the position. From the root-locus diagram, it can be determined that a proportional controller (P) is sufficient to shift all closed-loop poles to stability. Using the diagram, the resulting closed-loop behavior, i.e.\ oscillation for an underdamped system, can be tuned visually. Plotting the Nyquist diagram (Fig. \ref{fig:edubal_siso_nyquist}), students can identify the stability margins and the critical controller gain. Finally the students implement the designed controller in a target model. The simulated characteristics are then validated experimentally on the actual robot.

\subsection{MIMO Full state-feedback control}
\label{sec:mimo_control}
As a task for students from advanced control theory, EduBal can directly be stabilized using a MIMO controller. The goal is to stabilize all four states of the system in eq. \eqref{eq:state_space} around the operating point $\bm{x}_0=\left(0,0,0,0\right)^T$. This controller concept may also serve as a performance baseline for other controllers. After deriving the linear model of the robot and implementing a wheel velocity controller, students are provided with the model parameters. The design of a full-state feedback controller then follows the same procedure as shown in Section \ref{sec:siso_pos_control}. With $\bm{Q}=\text{diag}\left(
10^0, 10^{-2}, 10^0, 10^{-2}\right)$ and $R=10^{2}$ the optimization of equation \eqref{eq:lqr_cost} gives the optimal controller parameters in continuous time:
\begin{equation}
\bm{K}_{\mathrm{LQR},4\times 4} = \left(-0.1, -2.04, -90.23, -14.97\right)
\end{equation}
As the closed-loop system matrix $\bm{A}_K=\left(\bm{A}-\bm{BK}\right)$ only posesses eigenvalues with negative real part $\text{Re}(s_i) < 0$ the controlled dynamics are stable.

\begin{figure}
\includegraphics[scale=1]{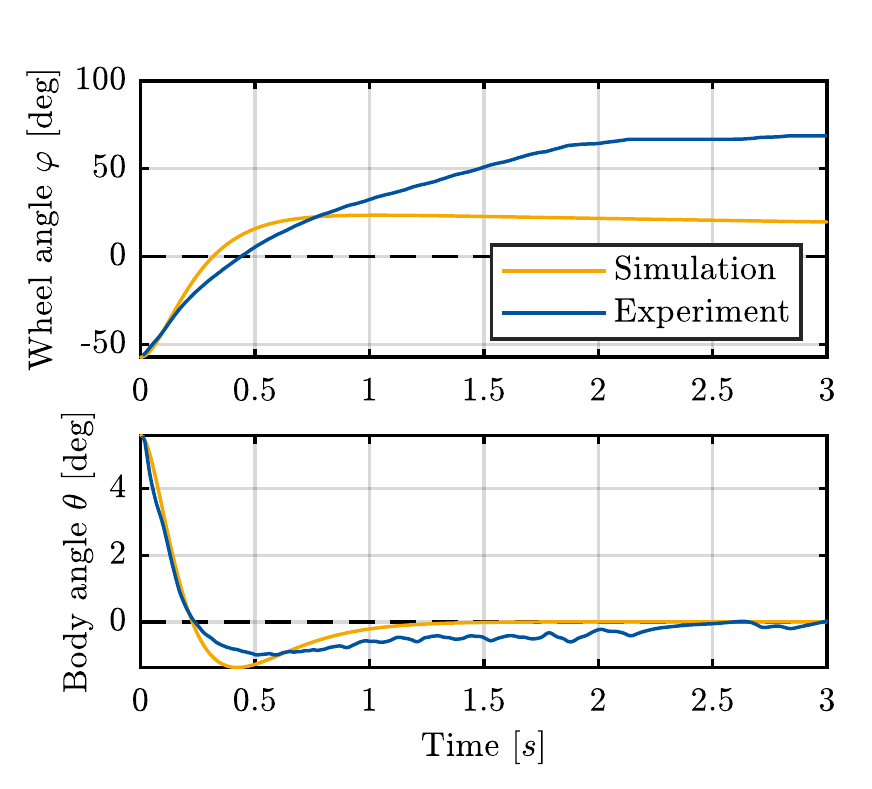}
\caption{Example trajectory of stabilized robot with full state-feedback control in simulation and experiment}
\label{fig:lqr_simulation_measured}
\end{figure}

Like in Section \ref{sec:siso_pos_control}, students are now able to validate the model quality and evaluate the controller performance on the actual robot. Fig. \ref{fig:lqr_simulation_measured} shows the closed-loop behavior for an initial state $\bm{x}_0=(\SI{-55}{\degree}, 0, \SI{5}{\degree}, 0)$ from simulation and experiment. Observing the behavior for $\varphi$ it can be noted that the robot successfully recovers from the initial disturbance $\varphi_0=\SI{-55}{\degree}$ but, especially in experiment, shows a large overshoot.  An explanation may be found in the unconsidered friction effects or in a slightly uneven ground during the experiment. The trajectory of $\theta$ shows similarities between simulation and experiment. The slight oscillations can be attributed to the gearbox backlash which is not included in the simulation model.
\section{Implementation}
\label{sec:implementation}
\subsection{Hardware design}
The EduBal robot is displayed in Fig. \ref{fig:edubal_overview}. It is assembled from a body, a stack of circuit boards, a battery and two electric motors. The complete robot weighs \SI{0.99}{\kg} and has the outer dimensions of $206\times 101 \times 235\si{\mm}$ ($W \times L \times H$). It is small enough to be placed and operated on a table. Bumpers are installed on the body to protect the circuit boards from damage in case of a fall. The body parts are made from PLA and are fully 3D-printable on a hobbyist machine. This allows for quick modification and replacement of individual parts. The \SI{9.9}{\volt} \SI{2100}{\milli\A\hour} \SI{40}{\coulomb} LiFePo battery is purposely placed high above the ground in a cage in order to raise the center of mass. As can be seen in Fig. \ref{fig:edubal_definitions} this asymmetric design in the lateral plane leads to an equilibrium body tilt angle of about \SI{10.5}{\deg} which is taken as controller reference.

Inside of the body is contained the circuit board stack with the Texas Instruments \emph{Launchpad XL} evaluation board. The F28069M microcontroller is based on the C2000-architecture and runs at \SI{90}{\mega\hertz}. Using the C2000 Embedded Coder Support Package, it is directly programmable via a USB cable. Contained in the support package is a blockset to access the low-level hardware interfaces. The board is extended with three custom circuit boards that are stacked on top and bottom of the microcontroller headers. The \emph{Power Supply Board} provides a \SI{3.3}{\volt} (\SI{2}{\A} maximum) and \SI{12}{\volt} (\SI{5}{\A} continuous) line from the nominal \SI{9.9}{\volt} battery. Given a PWM reference, the \emph{Motor Control Board} can directionally drive two \SI{12}{\volt} brushed DC electric motors using two H-bridges. For the EduBal two motors with an included gear transmission ratio of 30:1 are used. Each can deliver a wheel torque of \SI{0.18}{\N\m} at a longitudinal velocity of \SI{2.34}{\m\per\s} at maximum efficiency. The shaft angle of the motor is directly measured using a magnetic encoder. The third \emph{Logic board} combines a 9-axis inertial measurement unit (IMU) for measuring acceleration, angular rate and magnetic field, a Wi-Fi chip, user interaction elements and SPI and I2C headers for hardware extensions. All necessary robot components are compiled in a list which is available at the repository together with the CAD designs and PCB schematics. At the time of writing the cost of the necessary hardware for one unit is approximately \euro{200}.
\begin{figure*}
  \centering
  \includegraphics[width=\textwidth]{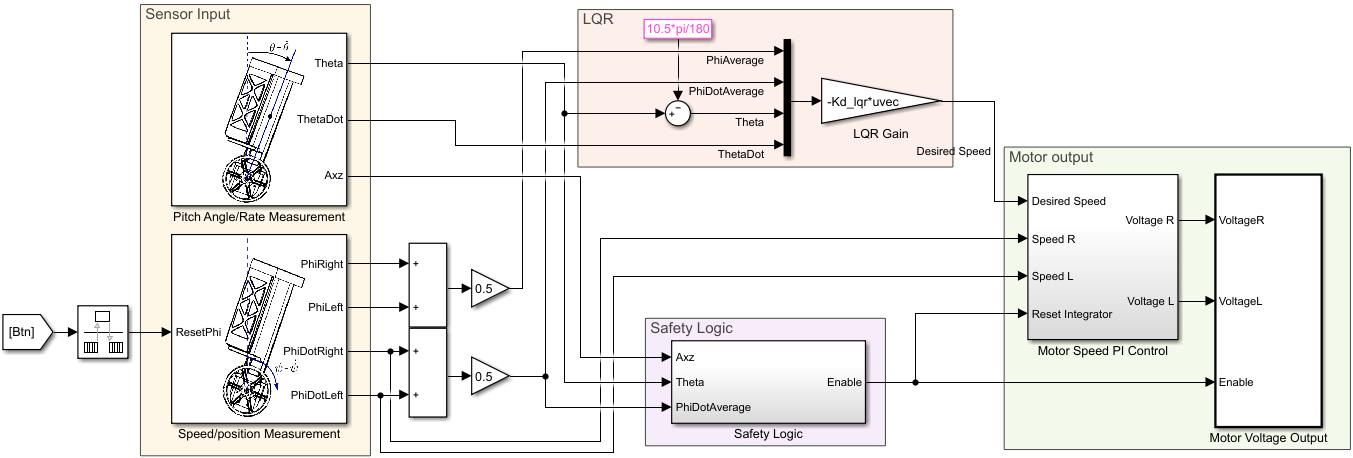}
  \caption{Implementation of the full state-feedback controller as explained in Section \ref{sec:mimo_control} using blocks from the provided Simulink library}
  \label{fig:mimo_control}
\end{figure*}
\subsection{Interfaces}
For direct interaction with the running model on the robot, it features four buttons, three LEDs and a buzzer on the front that are freely programmable. Using an attached USB cable, Simulink models can be run in External mode. In this mode, students can visualize live data in a Simulink Scope and change model parameters online. For remote interaction the onboard ESP8266 Wi-Fi chip can be configured to create an access point or join an existing network. It is connected to the microcontroller's serial port and converts the incoming and outgoing data stream to UDP packages. Depending on the Simulink model implementation, arbitrary data can be sent and received wirelessly. For demonstration purposes, we implemented wireless keyboard control and live-plotting of sensor signals in a MATLAB GUI. Depending on the application, EduBal can be extended with additional sensors and processing hardware. New 3D-printed parts can be attached to the chassis and the \emph{Logic Board} offers access to I2C and SPI headers to allow communication with the microcontroller board. As an example we designed a vision module and used it with EduBal (Fig. \ref{fig:line_follower}).

\subsection{Software}
Part of the EduBal repository is a software framework implemented in MATLAB/Simulink. It provides  tools to set up the hardware, i.e.\ an IMU calibration and a DC motor identification procedure. An analytical model of the robot dynamics derived in Section \ref{sec:model} is included alongside a respective simulation. In order to make the robot operational, a reference Simulink model is contained that offers the possibility to choose from different controllers and allows remote control operation. All models are built using blocks from a provided Simulink library. Included are blocks for accessing hardware interfaces such as the IMU and encoder input, motor output, user interaction elements and Wi-Fi communication. For angle estimation, a complementary filter and a Kalman filter implementation are available. An example implementation of the full state-feedback controller from Section \ref{sec:mimo_control} that uses the provided blocks is shown in Fig. \ref{fig:mimo_control}.
\subsection{Safety logic}
\begin{figure}
\centering
\graphicspath{{img/}}
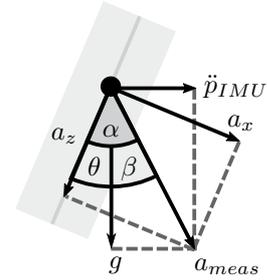
\caption{Coordinate transformation from sensor to global frame to estimate longitudinal acceleration from IMU measurements}
\label{fig:edubal_slip_imu}
\end{figure}

As the robot is used by students it must operate safely under all circumstances. For that purpose a few safety functions are implemented directly in the Simulink model. Consequently the model must run for them to be active.
\begin{itemize}
\item \emph{Battery monitor}: Using the balancer cable of the LiFePo battery, the individual cell voltages are monitored. In case of low voltage an audible warning is given to prevent battery draining damage.
\item \emph{Angle limiter}: In order to prevent unstable behavior for poor controller implementations, motor control is deactivated for a body angle exceeding the range of $\SI{-30}{\degree}<\theta<\SI{40}{\degree}$.
\item \emph{Wheel slip monitor}: Due to its small size, students tend to pick up the robot during operation. Depending on the controller implementation this leads to a rapid acceleration of the wheels. Putting the robot back down in this condition may cause unsafe behavior. To detect when the robot is picked up, we monitor the offset 
\begin{equation}
\Delta\ddot{p}=\ddot{p}_{\mathrm{IMU}}-\ddot{p}_{\mathrm{Whl}}
\end{equation}
where $\ddot{p}_{\mathrm{IMU}}$ is the longitudinal acceleration as measured by the IMU and $\ddot{p}_{\mathrm{Whl}}=r\ddot{\varphi}$ is the expected longitudinal acceleration from observed wheel acceleration. When the robot is picked up or when it drives on slippery ground, the resulting wheel slip is detectable as a peak in $\Delta p$. Fig. \ref{fig:edubal_slip_imu} shows the IMU attached to the tilted body. To calculate longitudinal acceleration $\ddot{p}_{\mathrm{IMU}}$ the measured accelerations $a_x, a_z$ in the sensor frame are transformed to the global frame using the estimated pitch angle $\theta$:
\begin{eqnarray}
a_{\mathrm{meas}} = \sqrt{a_x^2+a_z^2}\\
\alpha=\arctan \left(\frac{a_x}{a_z}\right)\\
\beta=\alpha - \theta\\
\ddot{p}_{\mathrm{IMU}} = \sin(\beta)\cdot a_{\mathrm{meas}}
\end{eqnarray}
We purposely neglect the small influence of the rotational acceleration $\ddot{\theta}$ on $\ddot{p}_{\mathrm{IMU}}$ as it can only be estimated by a discrete-time derivative and is thus susceptible to noise. 
\end{itemize}

\section{Educational use}
\label{sec:educational_use}
\begin{figure}
  \centering
  \includegraphics[width=0.5\columnwidth]{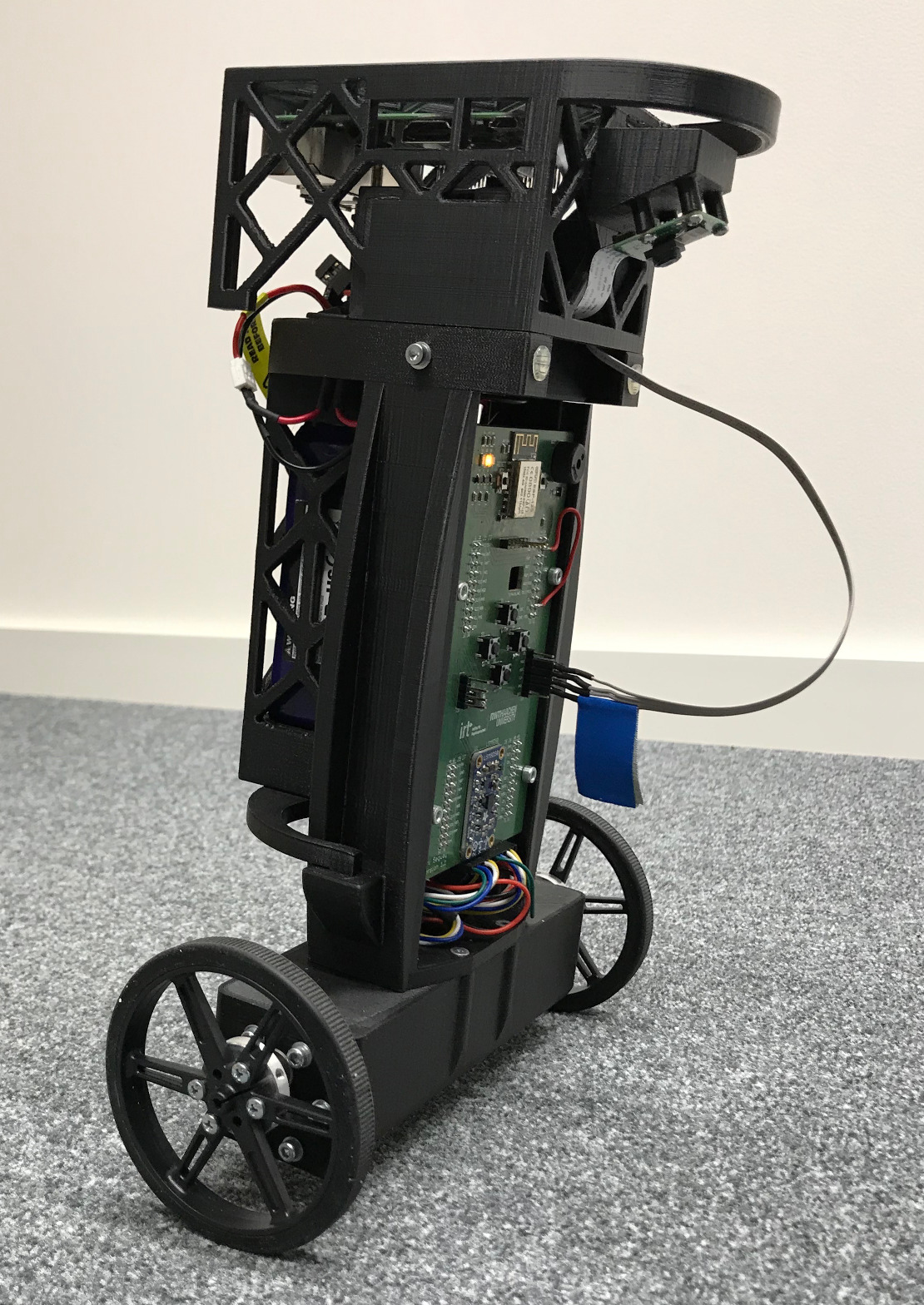}
  \caption{EduBal equipped with a vision module consisting of a Raspberry Pi 3B and camera used in the task of following a line on the ground}
  \label{fig:line_follower}
\end{figure}
The motivation behind EduBal is for it to serve as an example system for a variety of control tasks. Sections \ref{sec:electric_motor} and \ref{sec:mimo_control} are examples for tasks from system identification and control design that were given to RWTH Master students as part of a lecture about methods of Rapid Control Prototyping \citep{abelRapidControlPrototyping2006}. In the laboratory, 30 students in groups of three worked together in solving the tasks and implemented their controllers both in simulation and on the actual robot using the MATLAB/Simulink package. The Institute of Automatic Control developed the EduBal robot to replace a former human-sized balancing robot demonstrator. When introduced with the EduBal robots, students showed high curiosity to learn about the involved hardware. Provided with instructions, they were able to handle the robot without close supervision. The system was interactively analyzed and the controllers tuned using Simulink's External mode. After solving the given tasks students quickly started a discussion about new creative ways of applying concepts from advanced control-theoretic lectures to this system.

At the Institute for Dynamic Systems and Control of ETH Zurich, EduBal was adopted and it is planned to be used as part of both undergraduate and graduate lectures in control theory. In the exercises students derive a linear model of the robot and analyze its stability. The position control problem from Section \ref{sec:siso_pos_control} was developed to analyze transient system behavior of the stabilized robot as well as to design a SISO controller using the root-locus method. Students are also tasked to examine the effects of parameter uncertainty and gearbox backlash nonlinearity. As the Bachelor's course involves a number of 420 students, 20 robots were built beforehand.

Apart from the use in coursework, EduBal has been the subject of research to individual students. In order to improve robust behavior for varying system parameters, the implementations of a Sliding Mode and a $H_{\infty}$ controller were evaluated. With the goal of automatically following a line on the ground, the robot was extended with a vision module shown in Fig. \ref{fig:line_follower}. Based on the camera input a student team implemented a line detection algorithm and a lateral guidance controller that provides a wheel velocity reference to the EduBal onboard controller.

\section{Conclusion}
\label{sec:conclusion}
In this work we presented EduBal, an open-source design of a balancing robot aiming for the interactive education of control theory. The robot is built from 3D printed parts and custom circuit boards and due to its low cost can be built in greater numbers for the use in laboratories. Students can fully program the robot via Simulink using the provided block library for interfacing with sensor inputs and motor output. As an educational tool, it allows teaching a wide area of control theory in practice of which example tasks from system identification, SISO and MIMO control were provided.

Current work aims for the integration of the robot into advanced control curriculum. For that purpose the implementation of (Nonlinear) Model Predictive Control is investigated. Targeting the automatic parametrization of the available controllers in experiment, safe bayesian optimization is examined with the robot.

%%%%%%%%%%%%%%%%%%%%%%%%%%%%%%%%%%%%%%%%%%%%%%%%%%%%%%%%%%%%%%%%%%%%%%%%%%%%%%%%
%\tiny
\bibliography{literature}
\end{document}

%% file: img/edubal_overview.pdf_tex
%% Accompanies image file 'edubal_overview.pdf' (pdf, eps, ps)
\begingroup%
  \makeatletter%
  \providecommand\color[2][]{%
    \errmessage{(Inkscape) Color is used for the text in Inkscape, but the package 'color.sty' is not loaded}%
    \renewcommand\color[2][]{}%
  }%
  \providecommand\transparent[1]{%
    \errmessage{(Inkscape) Transparency is used (non-zero) for the text in Inkscape, but the package 'transparent.sty' is not loaded}%
    \renewcommand\transparent[1]{}%
  }%
  \providecommand\rotatebox[2]{#2}%
  \newcommand*\fsize{\dimexpr\f@size pt\relax}%
  \newcommand*\lineheight[1]{\fontsize{\fsize}{#1\fsize}\selectfont}%
  \ifx\svgwidth\undefined%
    \setlength{\unitlength}{451.95181201bp}%
    \ifx\svgscale\undefined%
      \relax%
    \else%
      \setlength{\unitlength}{\unitlength * \real{\svgscale}}%
    \fi%
  \else%
    \setlength{\unitlength}{\svgwidth}%
  \fi%
  \global\let\svgwidth\undefined%
  \global\let\svgscale\undefined%
  \makeatother%
  \begin{picture}(1,0.94661623)%
    \lineheight{1}%
    \setlength\tabcolsep{0pt}%
    \put(0,0){\includegraphics[width=\unitlength,page=1]{edubal_overview.pdf}}%
    \put(-0.00000085,0.64540805){\color[rgb]{0,0,0}\makebox(0,0)[lt]{\begin{minipage}{0.33526336\unitlength}\raggedright Launchpad XL Power supply, Motor controller, Logic board\end{minipage}}}%
    \put(0.67220217,0.15424723){\color[rgb]{0,0,0}\makebox(0,0)[lt]{\begin{minipage}{0.33526336\unitlength}\raggedright 2$\times$\SI{12}{\volt} Brushed DC motors\end{minipage}}}%
    \put(-0.00000085,0.91194242){\color[rgb]{0,0,0}\makebox(0,0)[lt]{\begin{minipage}{0.33526336\unitlength}\raggedright \SI{9.9}{\volt} LiFePo battery\end{minipage}}}%
  \end{picture}%
\endgroup%

%% file: img/edubal_definitions.pdf_tex
%% Accompanies image file 'edubal_definitions.pdf' (pdf, eps, ps)
\begingroup%
  \makeatletter%
  \providecommand\color[2][]{%
    \errmessage{(Inkscape) Color is used for the text in Inkscape, but the package 'color.sty' is not loaded}%
    \renewcommand\color[2][]{}%
  }%
  \providecommand\transparent[1]{%
    \errmessage{(Inkscape) Transparency is used (non-zero) for the text in Inkscape, but the package 'transparent.sty' is not loaded}%
    \renewcommand\transparent[1]{}%
  }%
  \providecommand\rotatebox[2]{#2}%
  \newcommand*\fsize{\dimexpr\f@size pt\relax}%
  \newcommand*\lineheight[1]{\fontsize{\fsize}{#1\fsize}\selectfont}%
  \ifx\svgwidth\undefined%
    \setlength{\unitlength}{128.947828bp}%
    \ifx\svgscale\undefined%
      \relax%
    \else%
      \setlength{\unitlength}{\unitlength * \real{\svgscale}}%
    \fi%
  \else%
    \setlength{\unitlength}{\svgwidth}%
  \fi%
  \global\let\svgwidth\undefined%
  \global\let\svgscale\undefined%
  \makeatother%
  \begin{picture}(1,1.61753415)%
    \lineheight{1}%
    \setlength\tabcolsep{0pt}%
    \put(0,0){\includegraphics[width=\unitlength,page=1]{edubal_definitions.pdf}}%
    \put(0.05768597,1.32031005){\color[rgb]{0,0,0}\makebox(0,0)[lt]{\lineheight{1.25}\smash{\begin{tabular}[t]{l}$g$\end{tabular}}}}%
    \put(0,0){\includegraphics[width=\unitlength,page=2]{edubal_definitions.pdf}}%
    \put(0.04614425,0.83144296){\color[rgb]{0,0,0}\makebox(0,0)[lt]{\lineheight{1.25}\smash{\begin{tabular}[t]{l}$l$\end{tabular}}}}%
    \put(0,0){\includegraphics[width=\unitlength,page=3]{edubal_definitions.pdf}}%
    \put(0.10639928,0.19436705){\color[rgb]{0,0,0}\makebox(0,0)[lt]{\lineheight{1.25}\smash{\begin{tabular}[t]{l}$r$\end{tabular}}}}%
    \put(0,0){\includegraphics[width=\unitlength,page=4]{edubal_definitions.pdf}}%
    \put(0.81909969,1.05150946){\color[rgb]{0,0,0}\makebox(0,0)[lt]{\lineheight{1.25}\smash{\begin{tabular}[t]{l}$m$\end{tabular}}}}%
    \put(0,0){\includegraphics[width=\unitlength,page=5]{edubal_definitions.pdf}}%
    \put(0.50731021,0.80525137){\color[rgb]{0,0,0}\makebox(0,0)[lt]{\lineheight{1.25}\smash{\begin{tabular}[t]{l}$\theta$\end{tabular}}}}%
    \put(0.54155131,0.38077938){\color[rgb]{0,0,0}\makebox(0,0)[lt]{\lineheight{1.25}\smash{\begin{tabular}[t]{l}$\varphi$\end{tabular}}}}%
    \put(0,0){\includegraphics[width=\unitlength,page=6]{edubal_definitions.pdf}}%
    \put(0.77468057,0.10387225){\color[rgb]{0,0,0}\makebox(0,0)[lt]{\lineheight{1.25}\smash{\begin{tabular}[t]{l}$p$\end{tabular}}}}%
    \put(0.74309061,0.4588107){\color[rgb]{0,0,0}\makebox(0,0)[lt]{\lineheight{1.25}\smash{\begin{tabular}[t]{l}$T$\end{tabular}}}}%
    \put(0,0){\includegraphics[width=\unitlength,page=7]{edubal_definitions.pdf}}%
  \end{picture}%
\endgroup%

%% file: img/parameters_table.tex
\begin{table}
\centering
\caption{Model parameters as measured in experiment}
\begin{tabular}[pos]{l|c|c}
Parameter & Symbol & Value\\
\hline
Body mass & $m$ & \SI{9.33e-1}{\kilogram}\\
%Body inertia & $I$ & \SI{6.86e-3}{\kilogram\square\metre}\\
Wheel radius & $r$ & \SI{0.04}{\metre}\\
COM above wheel axle & $l$ & \SI{8.57e-2}{\metre}
\end{tabular}
\label{tbl:parameters_table}
\end{table}

%% file: img/edubal_siso.pdf_tex
%% Accompanies image file 'edubal_siso.pdf' (pdf, eps, ps)
\begingroup%
  \makeatletter%
  \providecommand\color[2][]{%
    \errmessage{(Inkscape) Color is used for the text in Inkscape, but the package 'color.sty' is not loaded}%
    \renewcommand\color[2][]{}%
  }%
  \providecommand\transparent[1]{%
    \errmessage{(Inkscape) Transparency is used (non-zero) for the text in Inkscape, but the package 'transparent.sty' is not loaded}%
    \renewcommand\transparent[1]{}%
  }%
  \providecommand\rotatebox[2]{#2}%
  \newcommand*\fsize{\dimexpr\f@size pt\relax}%
  \newcommand*\lineheight[1]{\fontsize{\fsize}{#1\fsize}\selectfont}%
  \ifx\svgwidth\undefined%
    \setlength{\unitlength}{246.80794206bp}%
    \ifx\svgscale\undefined%
      \relax%
    \else%
      \setlength{\unitlength}{\unitlength * \real{\svgscale}}%
    \fi%
  \else%
    \setlength{\unitlength}{\svgwidth}%
  \fi%
  \global\let\svgwidth\undefined%
  \global\let\svgscale\undefined%
  \makeatother%
  \begin{picture}(1,0.3650979)%
    \lineheight{1}%
    \setlength\tabcolsep{0pt}%
    \put(0,0){\includegraphics[width=\unitlength,page=1]{edubal_siso.pdf}}%
    \put(0.67414151,0.14797675){\color[rgb]{0,0,0}\makebox(0,0)[lt]{\begin{minipage}{0.22749822\unitlength}\centering EduBal\end{minipage}}}%
    \put(0,0){\includegraphics[width=\unitlength,page=2]{edubal_siso.pdf}}%
    \put(0.61495717,0.14937725){\color[rgb]{0,0,0}\makebox(0,0)[lt]{\lineheight{1.25}\smash{\begin{tabular}[t]{l}$\dot{\varphi}_{\mathrm{ref}}$\end{tabular}}}}%
    \put(0,0){\includegraphics[width=\unitlength,page=3]{edubal_siso.pdf}}%
    \put(0.46245614,0.11634096){\color[rgb]{0,0,0}\makebox(0,0)[lt]{\lineheight{1.25}\smash{\begin{tabular}[t]{l}$C_{\mathrm{MIMO}}$\end{tabular}}}}%
    \put(0,0){\includegraphics[width=\unitlength,page=4]{edubal_siso.pdf}}%
    \put(0.56826004,0.23517204){\color[rgb]{0,0,0}\makebox(0,0)[lt]{\lineheight{1.25}\smash{\begin{tabular}[t]{l}$\bm{x}$\end{tabular}}}}%
    \put(0,0){\includegraphics[width=\unitlength,page=5]{edubal_siso.pdf}}%
    \put(0.16383998,0.10456478){\color[rgb]{0,0,0}\makebox(0,0)[lt]{\lineheight{1.25}\smash{\begin{tabular}[t]{l}$C_{\mathrm{SISO}}$\end{tabular}}}}%
    \put(0,0){\includegraphics[width=\unitlength,page=6]{edubal_siso.pdf}}%
    \put(0.10662417,0.07095628){\color[rgb]{0,0,0}\makebox(0,0)[lt]{\lineheight{1.25}\smash{\begin{tabular}[t]{l}$-$\end{tabular}}}}%
    \put(0.04757836,0.07095628){\color[rgb]{0,0,0}\makebox(0,0)[lt]{\lineheight{1.25}\smash{\begin{tabular}[t]{l}$+$\end{tabular}}}}%
    \put(-0.00054043,0.14440364){\color[rgb]{0,0,0}\makebox(0,0)[lt]{\lineheight{1.25}\smash{\begin{tabular}[t]{l}$p_{\mathrm{ref}}$\end{tabular}}}}%
    \put(0,0){\includegraphics[width=\unitlength,page=7]{edubal_siso.pdf}}%
    \put(0.92667388,0.14937728){\color[rgb]{0,0,0}\makebox(0,0)[lt]{\lineheight{1.25}\smash{\begin{tabular}[t]{l}$p$\end{tabular}}}}%
    \put(0.38339252,0.27866308){\color[rgb]{0,0,0}\makebox(0,0)[lt]{\lineheight{1.25}\smash{\begin{tabular}[t]{l}Stabilized EduBal (SISO)\end{tabular}}}}%
    \put(0.57714367,0.33697823){\color[rgb]{0,0,0}\makebox(0,0)[lt]{\lineheight{1.25}\smash{\begin{tabular}[t]{l}$G(s)$\end{tabular}}}}%
    \put(0.29191652,0.13224842){\color[rgb]{0,0,0}\makebox(0,0)[lt]{\lineheight{1.25}\smash{\begin{tabular}[t]{l}$\theta_{\mathrm{ref}}$\end{tabular}}}}%
  \end{picture}%
\endgroup%

%% file: img/edubal_slip_imu.pdf_tex
%% Accompanies image file 'edubal_slip_imu.pdf' (pdf, eps, ps)
\begingroup%
  \makeatletter%
  \providecommand\color[2][]{%
    \errmessage{(Inkscape) Color is used for the text in Inkscape, but the package 'color.sty' is not loaded}%
    \renewcommand\color[2][]{}%
  }%
  \providecommand\transparent[1]{%
    \errmessage{(Inkscape) Transparency is used (non-zero) for the text in Inkscape, but the package 'transparent.sty' is not loaded}%
    \renewcommand\transparent[1]{}%
  }%
  \providecommand\rotatebox[2]{#2}%
  \newcommand*\fsize{\dimexpr\f@size pt\relax}%
  \newcommand*\lineheight[1]{\fontsize{\fsize}{#1\fsize}\selectfont}%
  \ifx\svgwidth\undefined%
    \setlength{\unitlength}{93.26635295bp}%
    \ifx\svgscale\undefined%
      \relax%
    \else%
      \setlength{\unitlength}{\unitlength * \real{\svgscale}}%
    \fi%
  \else%
    \setlength{\unitlength}{\svgwidth}%
  \fi%
  \global\let\svgwidth\undefined%
  \global\let\svgscale\undefined%
  \makeatother%
  \begin{picture}(1,1.10440226)%
    \lineheight{1}%
    \setlength\tabcolsep{0pt}%
    \put(0,0){\includegraphics[width=\unitlength,page=1]{edubal_slip_imu.pdf}}%
    \put(0.75382885,0.73573931){\color[rgb]{0,0,0}\makebox(0,0)[lt]{\lineheight{1.25}\smash{\begin{tabular}[t]{l}$\ddot{p}_{IMU}$\end{tabular}}}}%
    \put(0.29012639,0.40165166){\color[rgb]{0,0,0}\makebox(0,0)[lt]{\lineheight{1.25}\smash{\begin{tabular}[t]{l}$\theta$\end{tabular}}}}%
    \put(0,0){\includegraphics[width=\unitlength,page=2]{edubal_slip_imu.pdf}}%
    \put(0.37367823,0.02817336){\color[rgb]{0,0,0}\makebox(0,0)[lt]{\lineheight{1.25}\smash{\begin{tabular}[t]{l}$g$\end{tabular}}}}%
    \put(0.71579603,0.01928183){\color[rgb]{0,0,0}\makebox(0,0)[lt]{\lineheight{1.25}\smash{\begin{tabular}[t]{l}$a_{meas}$\end{tabular}}}}%
    \put(0,0){\includegraphics[width=\unitlength,page=3]{edubal_slip_imu.pdf}}%
    \put(0.42212748,0.40368253){\color[rgb]{0,0,0}\makebox(0,0)[lt]{\lineheight{1.25}\smash{\begin{tabular}[t]{l}$\beta$\end{tabular}}}}%
    \put(0.34901914,0.55802208){\color[rgb]{0,0,0}\makebox(0,0)[lt]{\lineheight{1.25}\smash{\begin{tabular}[t]{l}$\alpha$\end{tabular}}}}%
    \put(0,0){\includegraphics[width=\unitlength,page=4]{edubal_slip_imu.pdf}}%
    \put(0.14326616,0.54717157){\color[rgb]{0,0,0}\makebox(0,0)[lt]{\lineheight{1.25}\smash{\begin{tabular}[t]{l}$a_z$\end{tabular}}}}%
    \put(0.84794736,0.59794114){\color[rgb]{0,0,0}\makebox(0,0)[lt]{\lineheight{1.25}\smash{\begin{tabular}[t]{l}$a_x$\end{tabular}}}}%
  \end{picture}%
\endgroup%